# Spin-orbit-derived giant magnetoresistance in a layered magnetic semiconductor AgCrSe$_2$

Hidefumi Takahashi[1,2,3], Tomoki Akiba[2], Alex Hiro Mayo[1,2], Kazuto Akiba[4], Atsushi Miyake[4], Masashi Tokunaga[4], Hitoshi Mori[5], Ryotaro Arita[5,6], and Shintaro Ishiwata[1,2,3]

[1] *Division of Materials Physics and Center for Spintronics Research Network (CSRN), Graduate School of Engineering Science, Osaka University, Toyonaka, Osaka 560-8531, Japan*
[2] *Department of Applied Physics, The University of Tokyo, Tokyo 113-8656, Japan*
[3] *Spintronics Research Network Division, Institute for Open and Transdisciplinary Research Initiatives, Osaka University, Yamadaoka 2-1, Suita, Osaka, 565-0871, Japan*
[4] *The Institute for Solid State Physics, The University of Tokyo, Kashiwa, Chiba 277-8581, Japan*
[5] *RIKEN Center for Emergent Matter Science, 2-1 Hirosawa, Wako, 351-0198, Japan*
[6] *Research Center for Advanced Science and Technology, University of Tokyo, Tokyo 153-8904, Japan*

Two-dimensional magnetic materials have recently attracted great interest due to their unique functions as the electric field control of a magnetic phase and the anomalous spin Hall effect. For such remarkable functions, a spin-orbit coupling (SOC) serves as an essential ingredient. Here we report a giant positive magnetoresistance in a layered magnetic semiconductor AgCrSe$_2$, which is a manifestation of the subtle combination of the SOC and Zeeman-type spin splitting. When the carrier concentration, $n$, approaches the critical value of 2.5×10$^{18}$ cm$^{-3}$, a sizable positive magnetoresistance of ~400 % emerges upon the application of magnetic fields normal to the conducting layers. Based on the magneto-Seebeck effect and the first-principles calculations, the unconventional magnetoresistance is ascribable to the enhancement of effective carrier mass in the SOC induced $J = 3/2$ state, which is tuned to the Fermi level through the Zeeman splitting enhanced by the $p$-$d$ coupling. This study demonstrates a new aspect of the SOC-derived magnetotransport in two-dimensional magnetic semiconductors, paving the way to novel spintronic functions.

## I. INTRODUCTION

Magnetic semiconductors are a key class of materials in the field of spintronics because of their potential for spin and charge manipulation. The combination of magnetism and spin-orbit coupling (SOC) yields versatile properties for logic functionalities and information storage capabilities[1–5]. In particular, two-dimensional systems, including van der Waals magnets such as the transition metal chalcogenides and halides[6–13], are being actively investigated for novel magnetism and spintronic applications owing to their advantage of creating thin films and heterostructures by exfoliation[11,12].

One of the promising two-dimensional magnetic semiconductors for the novel functions is AgCrSe$_2$ with layered structure[14–16]. Depending on the slight off-stoichiometry of the samples, the electric conductivity changes largely from semiconducting to metallic[17,18]. It is known to show a helimagnetic transition at 45 K[19]. Here, we report a novel type of giant positive magnetoresistance in a two-dimensional magnetic semiconductor AgCrSe$_2$ with a finely tuned carrier density. Considering the first-principles calculations in addition to the observation of the carrier dependent magnetoresistance and magneto-Seebeck effect, the mechanism of the giant magnetoresistance is discussed in terms of the subtle combination of the spin-orbit coupling (SOC) and the magnetically enhanced Zeeman splitting.

AgCrSe$_2$ has a layered polar structure lacking inversion symmetry (space group: $R3m$) at room temperature. As shown in Fig. 1a, the edge-shared CrSe$_6$ octahedra form layered triangular lattices with accommodating Ag$^+$ ions at one of the energetically degenerate tetrahedral sites (α- and β-sites)[14,16], giving rise to the polarity along the $c$ axis. It has been reported that the Ag$^+$ ions are distributed randomly in the α and β sites above 450-475 K, where the system undergoes an order-to-disorder transition to be a centrosymmetric structure with $R$-$3m$.[15] The edge-shared CrSe$_6$ octahedral layers are weakly coupled by Ag$^+$ ions across the van der Waals gap.[16]

## II. METHODS

*Material Synthesis and Characterization.* Single crystals of AgCrSe$_2$ were grown by the chemical vapor transport method using CrCl$_3$ as a transport agent[17]. A mixture of Ag powder (99.9 % purity), chromium powder (99.9 % purity) and selenium powder (99.9 % purity) in a stoichiometric ratio of 1:1:2 was sealed in an evacuated silica tube. The ampoule was placed in a three-zone furnace with a typical temperature gradient from 830 °C to 800 °C. After 1 to 2 weeks, the sample was cooled down to room temperature. The typical dimensions of the crystals are 5 × 5 × 0.05 mm$^3$. Single-crystal X-ray diffraction data were collected using a Rigaku XtaLAB mini II diffractometer with graphite monochromated Mo Kα radiation.

*Transport Measurements.* Electrical resistivity and Seebeck coefficient under magnetic fields were measured using a Physical Property Measurement System (Quantum Design, Inc.). The Seebeck coefficient was measured by the steady-state method. The magnetization and resistivity up to 45 T were measured by the nondestructive pulsed magnet with a pulse duration of 36 ms at the International MegaGauss Science Laboratory in the Institute for Solid State Physics, University of Tokyo.

*Density Functional Theory Calculations.* The relativistic and non-relativistic bulk electronic structure calculations of $AgCrSe_2$ were performed within the density functional theory (DFT) using the Perdew-Burke-Ernzerhof (PBE) exchange-correlation functional as implemented in the Quantum ESPRESSO code.[20–23]. The projector augmented wave (PAW) method has been used to account for the treatment of core electrons[23]. To properly treat the strong on-site Coulomb interaction of Cr-3$d$ states, an effective Hubbard-like potential term $U_{eff}$ was added. The value of $U_{eff}$ was fixed at 3 eV for Cr-3$d$ orbitals and zero for all other orbitals. The cut-off energy for plane waves forming the basis set was set to 60 eV. The lattice parameters and atomic position were taken from the experiment. The corresponding Brillouin zone was sampled using a $8 \times 8 \times 8$ $k$-mesh. To stabilize each FM configuration, the initial magnetic moment of Cr was set to 3 $\mu_B$ and aligned along the $c$ axis.

## III. RESULTS AND DISCUSSION

Figure 1b shows the temperature dependence of the magnetic susceptibility under a magnetic field ($H$) of 1T parallel to the $c$ axis. An antiferromagnetic (helimagnetic) transition can be found at 45 K as reported previously[24]. The hysteresis between the field-cooling (FC) and the zero-field-cooling (ZFC) processes below 45 K implies the spin canting due to the polar lattice distortion. We calculated the electronic band structure for the induced ferromagnetic state with spins aligned along the $c$ axis with the presence of SOC as shown in Fig. 1c. The valence bands with parabolic dispersion are seen around Γ point with small energy gaps (direct ~ 0.7 eV and indirect ~ 0.1 eV). As shown in Fig. 1d, the temperature dependence of in-plane resistivity varies widely from insulating to metallic, depending on the samples. This presumably reflects the small energy gap and the presence of naturally introduced defects at the Ag site, which yields hole-type carriers as confirmed by the positive sign both in the Hall resistivity (See Fig. S1 in Supplementary Information [25]) and the Seebeck coefficient (See Figs. 4 and S2 in Supplementary Information [25]). For transport measurements in this work, we selected three types of single crystals, #1, #2, and #3, with different carrier concentration[18].

Figures 2a-c show the magnetoresistance MR [$\Delta\rho(H)/\rho(0) = [\rho(H)-\rho(0)]/\rho(0)$)] for the three samples under $H//c$. Interestingly, the magnetic field dependences of MR are distinct for each sample. The samples #1 and #3 show negative MR at low temperatures, reflecting the suppression of the spin scattering as typically found in a spin-polarized ferromagnetic state. It should be noted that a significant positive MR as large as 50 % at 2 K and 7 T can be found for #2 below 100 K. To get insight into the origin of the remarkable MR in #2, we plot the carrier concentration $n$ dependence of the MR at 9 T, as shown in Fig. 2d. $n$ for each sample at selected temperatures was estimated as $n = 1/eR_H$ ($n$ for #1, #2, and #3 at 2 K were estimated to be $1\times 10^{16}$ cm$^{-3}$, $4\times10^{18}$ cm$^{-3}$, and $7\times10^{20}$ cm$^{-3}$, respectively). Here $R_H$ was evaluated from the Hall resistivity measurements (For details, see Fig. S1 in Supplementary Information [25]). Surprisingly, the carrier concentration dependence exhibits a kind of the scaling behavior denoted by the grey line, and has a sharp peak around $n = 2.5\times10^{18}$ cm$^{-3}$, suggesting the existence of critical carrier concentration for the significant positive MR. One of the possible origins of positive MR is the Lorentz force [MR = $\Delta\rho(H)/\rho(0) \sim (\mu H)^2$, where $\mu$ denotes carrier mobility] typically observed in high-mobility semiconductors and semimetals[26,27]. However, this is unlikely to be the case for the present system, since $\mu$ (=$1/\rho ne$; $e$ is elementary charge) is as low as ~ 20 cm$^2$/Vs (#2). To observe positive MR as large as ~ 50 % at 9 T, $\mu$ should be in the order of 1000 cm$^2$/Vs.

In order to identify the role of the magnetism in the positive MR, we measured the magnetization and MR for #2 at high $H$ up to 45 T at several temperatures with the field configurations $H//c$ (Figs. 3a,b) and $H\perp c$ (Figs. 3c,d). It is worth noting that under $H//c$, the MR reaches a quite large value higher than 400 %, and has a peak at 25 T, where the magnetic moment is saturated ($H_{IF}$). Analogous behavior is also observed under $H \perp c$, while the maximum MR is relatively small (~ 200 %). Note that the kink-like behavior around 3 T in the magnetization and MR curves can be associated with the spin-flop transition ($H_{SF}$), reflecting the planar spin anisotropy. These results strongly suggest the strong coupling between the localized moments and itinerant electrons.

Next, we measured magneto-Seebeck effect, which is sensitive to the change in the electronic state near the Fermi level, for the samples #2 and #3 under $H//c$ (the resistivity of #1 is too high to measure Seebeck coefficient). Figures 4a and 4b show the temperature dependence of Seebeck coefficient $S$ at $H = 0$ T and 9 T. While the Seebeck coefficient of both samples decreases with decreasing the temperature, similar to the heavily-doped (degenerate) semiconductors, only #2 underwent a substantial enhancement of $S$ under the application of magnetic fields. Figures 4c and 4d depict the magnetic field dependence of the magneto-Seebeck [MS=$S(H)$-$S(0)$] effect, which remarkably increases by the application of magnetic fields only for #2 with positive MR, implying that the same origin plays an important role on the significant positive MR and MS. Such a large enhancement of MS is not common in conventional metals or semiconductors[28], and the application of external magnetic fields on magnetic materials tends to reduce $S$ through the suppression of the magnetic entropy.[29,30] Thus, the significant positive MS in $AgCrSe_2$ is unusual and it is reasonable to consider the spin-driven band modification or reconstruction as its origin.[31]

Based on the Mott formula typically used for metallic systems, Seebeck coefficient is described as

$$S = (\pi^2 k_B^2 T m^*/3e\hbar^2)(3\pi^2 n)^{-2/3}, \qquad (1)$$

where $m^*$ is an effective carrier mass. According to this formula, the positive MS can be associated with the variation of $n$ or $m^*$ as a function of $H$. On the other hand, the electrical resistivity considering the Drude model is given by $\rho=1/(ne\mu)=m^*/(ne^2\tau)$, where $\tau$ is the carrier scattering time. Here we assume that $n$ is constant upon the change in $H$, as supported by the linear Hall resistivity $\rho_{yx}$ as a function of $H$ (see Fig. S1 in Supplementary Information [25]). Thus, the $H$-dependence of $S/\rho T$ presumably depends only on $\tau$. For instance, as shown in Fig. 4f, the $S/\rho T$ increases as a function of $H$ at low temperatures, which implies that the negative MR for #3 is caused by the $H$-induced increase of $\tau$, reflecting the suppression of the spin scattering. On the other hand, $S/\rho T$ for #2 is almost constant as shown in Fig. 4e, indicating that $\tau$ is independent of $H$, i.e. the magnetoresistance and magneto-Seebeck effect normalized by the zero-field values show almost the same magnetic field dependence at selected temperatures, as shown in Fig. S3 (Supplementary Information [25]). Therefore, given the Drude model, it is reasonable to presume that the significant positive MR as well as MS observed for #2 is not attributable to the change in $\tau$ but to the increase in $m^*$ by the application of $H$.

As the probable model of the $H$-induced band modification, we consider a spin-orbit coupled Zeeman-type spin splitting for the paramagnetic (PM) and ferromagnetic (FM) phases [32–34]. Figure 5a shows the calculated band structure of the PM phase, where the 3$d$ orbitals of Cr are excluded to reflect the experimental fact that they open an energy gap and become localized orbitals[35]. Note that we adopt the PM phase suitable for the calculation as the zero-field phase instead of the antiferromagnetic phase. This is because the band structure around the Γ point is expected to dominate the charge transport and be immune to the band folding upon the antiferromagnetic transition. Consistent with the expectations, there is no apparent anomaly in the electrical resistivity and the Seebeck coefficient at the magnetic transition temperature as shown in Fig. 1d and Fig. S2. The valence band structure of AgCrSe$_2$ is mainly composed of the $p$-like Se bands with spin and orbital angular momentums, $S=1/2$ and $L=1$, respectively. For the PM phase with SOC, the valence bands are characterized by the total angular momentum $J=3/2$, where the outer (solid lines) and inner (broken lines) bands are $J_z=\pm 3/2$ and $J_z=\pm 1/2$, respectively. The similar bands characterized by $J$ have been discussed in typical semiconductors such as GaAs[36,37]. Here, the $J_z=\pm 3/2$ bands and the $J_z=\pm 1/2$ bands are termed a heavy hole (HH) with a heavy effective mass and a light hole (LH) with a light effective mass, respectively. Next, we show the band structure of the FM phase with SOC in Fig. 5b. The spin-split parabolic valence bands at Γ point are composed of $S_z=\pm 1/2$ and $L_z=\pm 1$, which are split through the $p$-$d$ exchange coupling. Here, the bands with spins parallel to the localized Cr spins ($S_z=+1/2$ bands) are plotted in red curves, and the bands with antiparallel spins ($S_z=-1/2$ bands) are plotted in blue curves.

To discuss the mechanism of the giant positive MR, we show the schematic models for the band structure around the Γ point for the PM and FM states with SOC (see Figs. 5c and 5d). In the PM state, the HH and LH bands are almost degenerate (The energy splitting between them is about 0.01 eV), and the effective mass of the holes near the Fermi level is averaged over. On the other hand, in the FM state, the degeneracy of HH and LH is lifted, resulting in the predominance of heavy-mass holes in HH, the effective mass of which is expected to be heavier than the average value in the PM state (see Fig. 5c and 5d). This situation corresponds to the case of #2 showing the enhancement of resistivity under magnetic fields. Namely, when the Fermi level $E_F$ lies near the band edge as in the case of #2 (Estimation of $E_F$ is discussed in the section S4 of Supplementary Information [25]), the transport properties are strongly influenced by this band modification caused by the combination of the SOC and the $p$-$d$ coupling. The importance of the SOC for this novel type of positive MR can be confirmed by the band structures for the FM state without SOC shown in Figs. S4 and S5 (Supplementary Information [25]). The $S_z=+1/2$ and $S_z=-1/2$ bands near the Fermi level emerging through the Zeeman splitting in the FM phase are expected to have similar dispersions as the spin-degenerated bands in the PM phase, because their orbital character remains intact upon the Zeeman splitting (see Figs. S5(b) and S5(c)). Therefore, the effective carrier mass is expected to be almost unchanged on the course of the field-induced transition from the PM phase to the FM phase. This explanation is consistent with the presence of the critical carrier concentration $n$ for the MR, and applicable also to the positive MR under in-plane magnetic fields ($H//ab$); the observed anisotropy of MR can be associated with the different field effect on the spin splitting of the HH bands (see Fig. S6 in Supplementary Information [25]).

## IV. CONCLUSION

In conclusion, we have observed giant positive MR in the layered magnetic semiconductor AgCrSe$_2$ through the fine tuning of the carrier concentration. The magneto-Seebeck effect and band calculations revealed that the novel MR effect can be ascribed to the significant enhancement of the effective carrier mass upon the spin splitting caused by the unique combination of the SOC and the $p$-$d$ coupling. This work demonstrates the importance of the SOC and the high carrier tunability in addition to the spin-charge coupling for the exploration of novel spintronic functions in magnetic two-dimensional materials.

### ACKNOWLEDGMENTS
The authors thank Y. Fuseya, A. Yamada, and T. Nomoto for fruitful discussions. This study was supported in part by KAKENHI (Grant No. 17H06137, 19H02424, 19K14660, 20K03802, and 21H01030), the Asahi Glass Foundation,

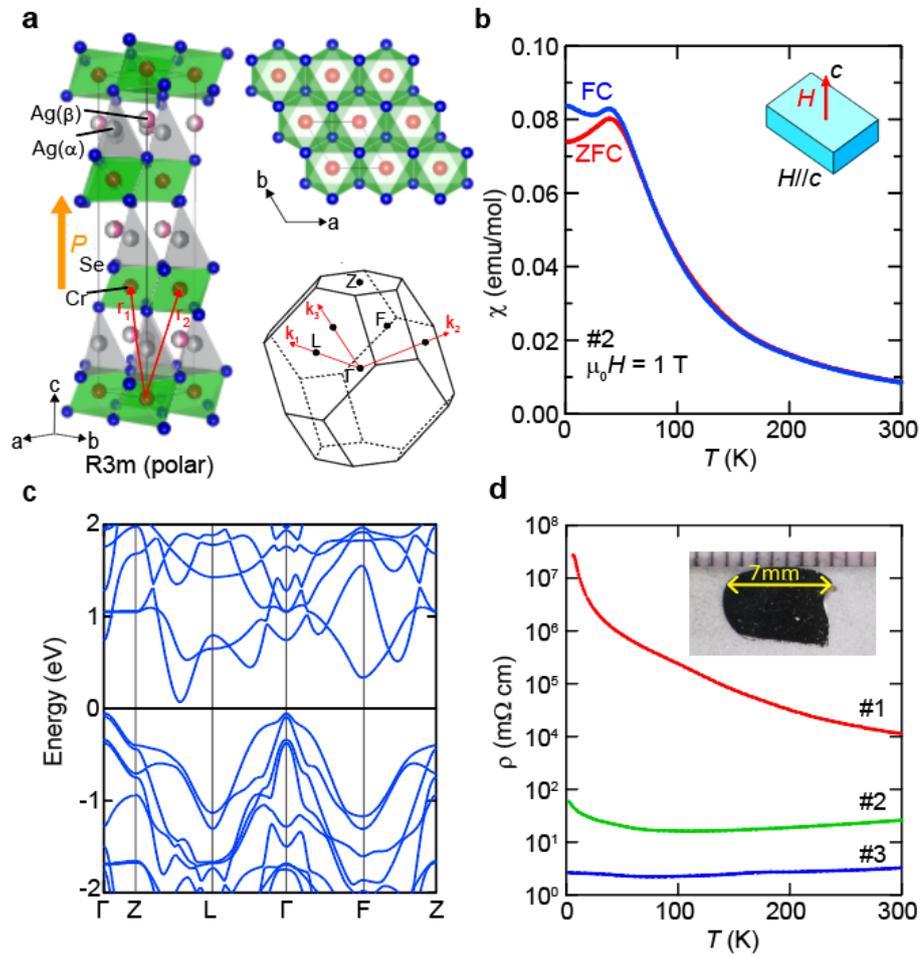

**FIG. 1.** (a) Crystal structure and Brillouin zone. Crystal structure is shown by the hexagonal cell (conventional cell). Brillouin zone is based on the rhombohedral cell (primitive cell). (b) Temperature dependence of magnetic susceptibility for #2 under out-of-plane magnetic field ($H = 1$ T). The overall behaviour is almost the same among the three samples (#1, #2, #3). (c) Electronic structure for the induced ferromagnetic state with spin-orbit coupling. Magnetic moments are parallel to $c$-axis. (d) Temperature dependence of electrical resistivity for three types of AgCrSe$_2$ crystals, denoted by #1, #2, and #3. Inset shows a typical single crystal.

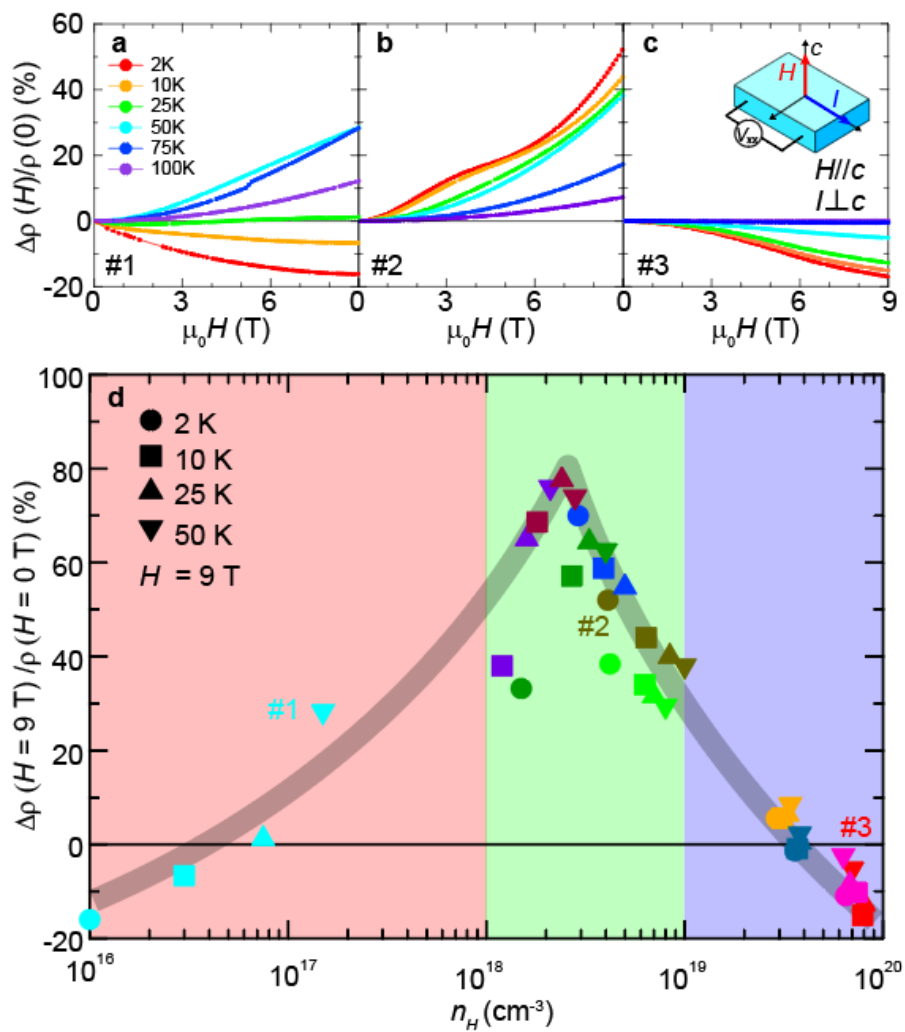

**FIG. 2.** Magnetoresistance in three types of AgCrSe$_2$ crystals: #1 (a), #2 (b), #3 (c). (d) Carrier concentration dependence of magnetoresistance at selected temperatures in various crystals.

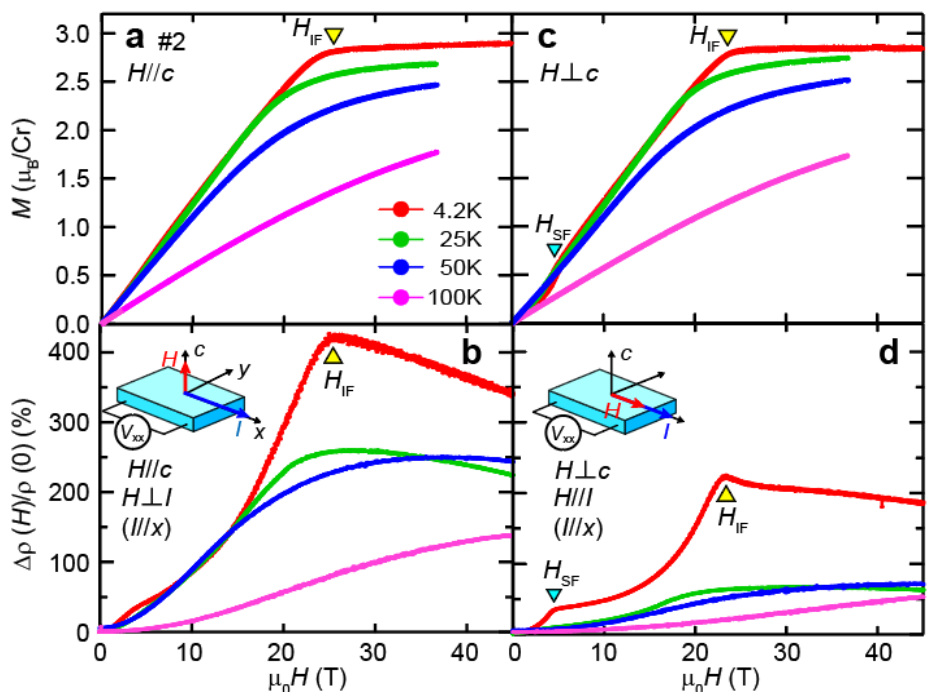

**FIG. 3.** Magnetic field $H$ dependence of (a,c) magnetization and (b,d) magnetoresistance under (a,b) out-of-plane magnetic field and, (c,d) in-plane magnetic field for sample #2.

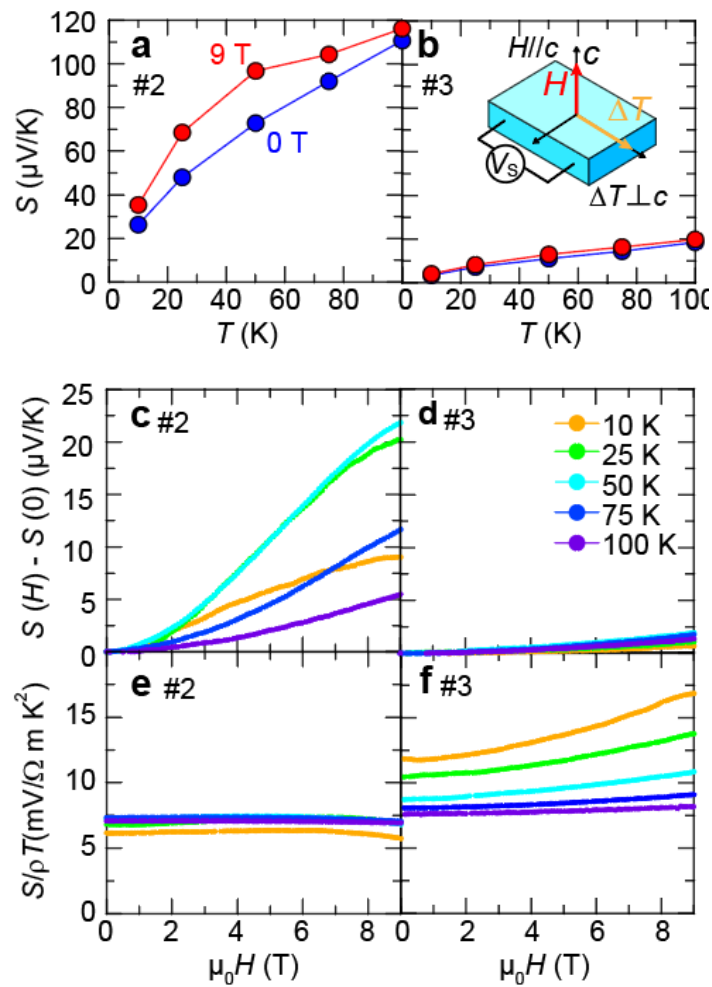

**FIG. 4.** Temperature dependence of magneto-Seebeck effect ($H//c$) in two types of AgCrSe$_2$ crystals: (a) #2, (b) #3. Magnetic field dependence of Seebeck effect for (c) #2 and (d) #3. Magnetic field dependence of $S/\rho T$ ($S$: Seebeck coefficient, $\rho$: electrical resistivity, $T$: temperature) for (e) #2 and (f) #3.

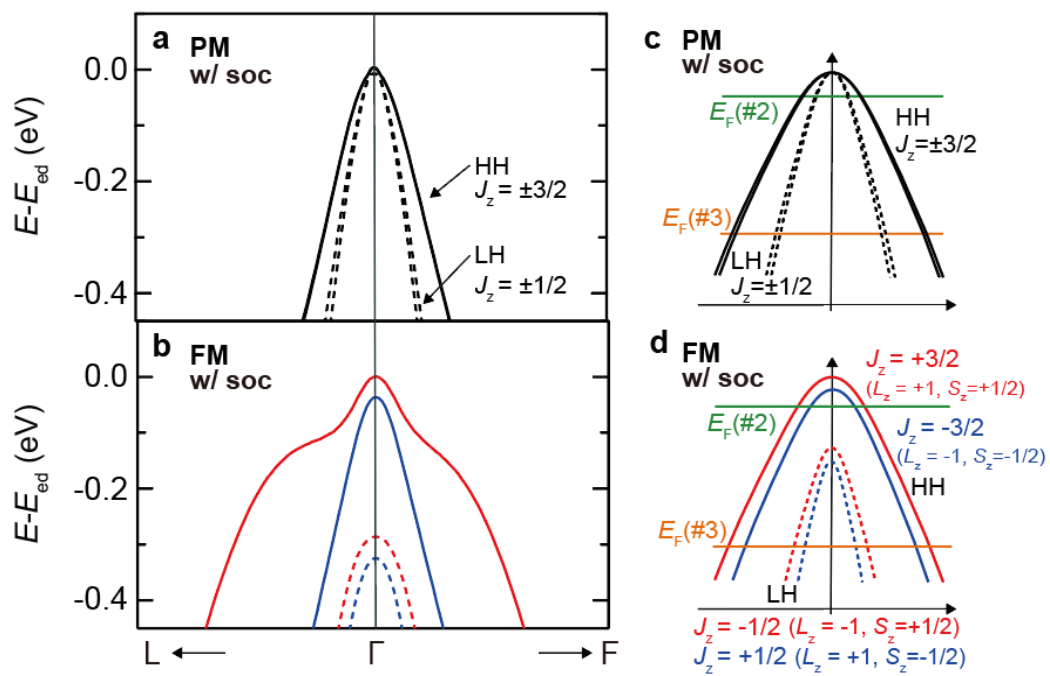

**FIG. 5.** Calculated b and structures for (a) the paramagnetic (PM) phase and (b) the ferromagnetic (FM) phase, respectively ($E_{ed}$ is the energy of the band edge at Γ point). The magnetic moments of Cr in the FM phase are aligned along the $c$ axis. Spin-orbit coupling is considered in the calculations. Schematic illustrations of the band structures corresponding to (c) the PM and (d) the FM phases, respectively.

# References

[1] X. Lin, W. Yang, K. L. Wang, and W. Zhao, Two-Dimensional Spintronics for Low-Power Electronics, Nat. Electron. **2**, 274 (2019).

[2] J. Grollier, D. Querlioz, K. Y. Camsari, K. Everschor-Sitte, S. Fukami, and M. D. Stiles, Neuromorphic Spintronics, Nat. Electron. **3**, 360 (2020).

[3] A. Hirohata, K. Yamada, Y. Nakatani, I.-L. Prejbeanu, B. Diény, P. Pirro, and B. Hillebrands, Review on Spintronics: Principles and Device Applications, J. Magn. Magn. Mater. **509**, 166711 (2020).

[4] H. Kurebayashi, J. H. Garcia, S. Khan, J. Sinova, and S. Roche, Magnetism, Symmetry and Spin Transport in van Der Waals Layered Systems, Nat. Rev. Phys. **4**, 150 (2022).

[5] M. Gibertini, M. Koperski, A. F. Morpurgo, and K. S. Novoselov, Magnetic 2D Materials and Heterostructures, Nat. Nanotechnol. **14**, 408 (2019).

[6] D. C. Freitas, R. Weht, A. Sulpice, G. Remenyi, P. Strobel, F. Gay, J. Marcus, and M. Núñez-Regueiro, Ferromagnetism in Layered Metastable 1T-$CrTe_2$, J. Phys. Condens. Matter **27**, 176002 (2015).

[7] D. C. Freitas, M. Núñez, P. Strobel, A. Sulpice, R. Weht, A. A. Aligia, and M. Núñez-Regueiro, Antiferromagnetism and Ferromagnetism in Layered 1T-$CrSe_2$ with V and Ti Replacements, Phys. Rev. B **87**, 014420 (2013).

[8] W.-B. Zhang, Q. Qu, P. Zhu, and C.-H. Lam, Robust Intrinsic Ferromagnetism and Half Semiconductivity in Stable Two-Dimensional Single-Layer Chromium Trihalides, J. Mater. Chem. **3**, 12457 (2015).

[9] M. A. McGuire, H. Dixit, V. R. Cooper, and B. C. Sales, Coupling of Crystal Structure and Magnetism in the Layered, Ferromagnetic Insulator $CrI_3$, Chem. Mater. **27**, 612 (2015).

[10] M. A. McGuire, G. Clark, S. Kc, W. M. Chance, G. E. Jellison, V. R. Cooper, X. Xu, and B. C. Sales, Magnetic Behavior and Spin-Lattice Coupling in Cleavable van Der Waals Layered $CrCl_3$ Crystals, Phys. Rev. Materials **1**, 014001 (2017).

[11] B. Huang, G. Clark, E. Navarro-Moratalla, D. R. Klein, R. Cheng, K. L. Seyler, D. Zhong, E. Schmidgall, M. A. McGuire, D. H. Cobden, W. Yao, D. Xiao, P. Jarillo-Herrero, and X. Xu, Layer-Dependent Ferromagnetism in a van Der Waals Crystal down to the Monolayer Limit, Nature **546**, 270 (2017).

[12] C. Gong, L. Li, Z. Li, H. Ji, A. Stern, Y. Xia, T. Cao, W. Bao, C. Wang, Y. Wang, Z. Q. Qiu, R. J. Cava, S. G. Louie, J. Xia, and X. Zhang, Discovery of Intrinsic Ferromagnetism in Two-Dimensional van Der Waals Crystals, Nature **546**, 265 (2017).

[13] W. Lin, K. Chen, S. Zhang, and C. L. Chien, Enhancement of Thermally Injected Spin Current through an Antiferromagnetic Insulator, Phys. Rev. Lett. **116**, 186601 (2016).

[14] D. W. Murphy, H. S. Chen, and B. Tell, Superionic Conduction in $AgCrS_2$ and $AgCrSe_2$, J. Electrochem. Soc. **124**, 1268 (1977).

[15] F. Gascoin and A. Maignan, Order–Disorder Transition in $AgCrSe_2$: A New Route to Efficient Thermoelectrics, Chem. Mater. **23**, 2510 (2011).

[16] B. Li, H. Wang, Y. Kawakita, Q. Zhang, M. Feygenson, H. L. Yu, D. Wu, K. Ohara, T. Kikuchi, K. Shibata, T. Yamada, X. K. Ning, Y. Chen, J. Q. He, D. Vaknin, R. Q. Wu, K. Nakajima, and M. G. Kanatzidis, Liquid-like Thermal Conduction in Intercalated Layered Crystalline Solids, Nat. Mater. **17**, 226 (2018).

[17] R. Yano and T. Sasagawa, Crystal Growth and Intrinsic Properties of $ACrX_2$ (A= Cu, Ag; X= S, Se) without a Secondary Phase, Cryst. Growth Des. **16**, 5618 (2016).

[18] Y. Shiomi, T. Akiba, H. Takahashi, and S. Ishiwata, Giant Piezoelectric Response in Superionic Polar Semiconductor, Adv. Electron. Mater. **4**, 1800174 (2018).

[19] F. M. R. Engelsman, G. A. Wiegers, F. Jellinek, and B. Van Laar, Crystal Structures and Magnetic Structures of Some metal(I) chromium(III) Sulfides and Selenides, J. Solid State Chem. **6**, 574 (1973).

[20] P. Giannozzi, S. Baroni, N. Bonini, M. Calandra, R. Car, C. Cavazzoni, D. Ceresoli, G. L. Chiarotti, M. Cococcioni, I. Dabo, A. Dal Corso, S. de Gironcoli, S. Fabris, G. Fratesi, R. Gebauer, U. Gerstmann, C. Gougoussis, A. Kokalj, M. Lazzeri, L. Martin-Samos, N. Marzari, F. Mauri, R. Mazzarello, S. Paolini, A. Pasquarello, L. Paulatto, C. Sbraccia, S. Scandolo, G. Sclauzero, A. P. Seitsonen, A. Smogunov, P. Umari, and R. M. Wentzcovitch, QUANTUM ESPRESSO: A Modular and Open-Source Software Project for Quantum Simulations of Materials, J. Phys. Condens. Matter **21**, 395502 (2009).

[21] P. Giannozzi, O. Andreussi, T. Brumme, O. Bunau, M. Buongiorno Nardelli, M. Calandra, R. Car, C. Cavazzoni, D. Ceresoli, M. Cococcioni, N. Colonna, I. Carnimeo, A. Dal Corso, S. de Gironcoli, P. Delugas, R. A. DiStasio Jr, A. Ferretti, A. Floris, G. Fratesi, G. Fugallo, R. Gebauer, U. Gerstmann, F. Giustino, T. Gorni, J. Jia, M. Kawamura, H.-Y. Ko, A. Kokalj, E. Küçükbenli, M. Lazzeri, M. Marsili, N. Marzari, F. Mauri, N. L. Nguyen, H.-V. Nguyen, A. Otero-de-la-Roza, L. Paulatto, S. Poncé, D. Rocca, R. Sabatini, B. Santra, M. Schlipf, A. P. Seitsonen, A. Smogunov, I. Timrov, T. Thonhauser, P. Umari, N. Vast, X. Wu, and S. Baroni, Advanced Capabilities for Materials Modelling with Quantum ESPRESSO, J. Phys. Condens. Matter **29**, 465901 (2017).

[22] Home Page, http://www.quantum-espresso.org.

[23] A. Dal Corso, Pseudopotentials Periodic Table: From H to Pu, Comput. Mater. Sci. **95**, 337 (2014).

[24] U. K. Gautam, R. Seshadri, S. Vasudevan, and A. Maignan, Magnetic and Transport Properties, and Electronic Structure of the Layered Chalcogenide $AgCrSe2$, Solid State Commun. **122**, 607 (2002).

[25] See Supplemental Material at ????? for more detailed information regarding Hall resistivity, Seebeck coefficient, estimation of Fermi energy, and band calculations.

[26] H. Takahashi, R. Okazaki, Y. Yasui, and I. Terasaki, Low-Temperature Magnetotransport of the Narrow-Gap Semiconductor $FeSb_2$, Phys. Rev. B **84**, 205215 (2011).

[27] M. N. Ali, J. Xiong, S. Flynn, J. Tao, Q. D. Gibson, L. M. Schoop, T. Liang, N. Haldolaarachchige, M. Hirschberger, N. P.

Ong, and R. J. Cava, Large, Non-Saturating Magnetoresistance in WTe2, Nature **514**, 205 (2014).
[28] J. M. Ziman, Principles of the Theory of Solids (Cambridge University Press, 1972).
[29] Y. Wang, N. S. Rogado, R. J. Cava, and N. P. Ong, Spin Entropy as the Likely Source of Enhanced Thermopower in $Na_{(x)}Co_2O_4$, Nature **423**, 425 (2003).
[30] T. D. Yamamoto, H. Taniguchi, Y. Yasui, S. Iguchi, T. Sasaki, and I. Terasaki, Magneto-Thermopower in the Weak Ferromagnetic Oxide CaRu0.8Sc0.2O3: An Experimental Test for the Kelvin Formula in a Magnetic Material, J. Phys. Soc. Jpn. **86**, 104707 (2017).
[31] K. Wang, D. Graf, and C. Petrovic, Large Magnetothermopower and Fermi Surface Reconstruction in $Sb_2Te_2Se$, Phys. Rev. B **89**, 125202 (2014).
[32] S. E. Barnes, J. 'ichi Ieda, and S. Maekawa, Rashba Spin-Orbit Anisotropy and the Electric Field Control of Magnetism, Sci. Rep. **4**, 4105 (2014).
[33] J. Krempaský, S. Muff, F. Bisti, M. Fanciulli, H. Volfová, A. P. Weber, N. Pilet, P. Warnicke, H. Ebert, J. Braun, F. Bertran, V. V. Volobuev, J. Minár, G. Springholz, J. H. Dil, and V. N. Strocov, Entanglement and Manipulation of the Magnetic and Spin–orbit Order in Multiferroic Rashba Semiconductors, Nat. Commun. **7**, 13071 (2016).
[34] I. I. Klimovskikh, A. M. Shikin, M. M. Otrokov, A. Ernst, I. P. Rusinov, O. E. Tereshchenko, V. A. Golyashov, J. Sánchez-Barriga, A. Y. Varykhalov, O. Rader, K. A. Kokh, and E. V. Chulkov, Giant Magnetic Band Gap in the Rashba-Split Surface State of Vanadium-Doped BiTeI: A Combined Photoemission and Ab Initio Study, Sci. Rep. **7**, 3353 (2017).
[35] M. Baenitz, M. M. Piva, S. Luther, J. Sichelschmidt, K. M. Ranjith, H. Dawczak-Dębicki, M. O. Ajeesh, S.-J. Kim, G. Siemann, C. Bigi, P. Manuel, D. Khalyavin, D. A. Sokolov, P. Mokhtari, H. Zhang, H. Yasuoka, P. D. C. King, G. Vinai, V. Polewczyk, P. Torelli, J. Wosnitza, U. Burkhardt, B. Schmidt, H. Rosner, S. Wirth, H. Kühne, M. Nicklas, and M. Schmidt, Planar Triangular S=3/2 Magnet $AgCrSe_2$: Magnetic Frustration, Short Range Correlations, and Field-Tuned Anisotropic Cycloidal Magnetic Order, Phys. Rev. B **104**, 134410 (2021).
[36] A. R. Hamilton, R. Danneau, O. Klochan, W. R. Clarke, A. P. Micolich, L. H. Ho, M. Y. Simmons, D. A. Ritchie, M. Pepper, K. Muraki, and Y. Hirayama, The 0.7 Anomaly in One-Dimensional Hole Quantum Wires, J. Phys. Condens. Matter **20**, 164205 (2008).
[37] R. Winkler, Spin-Orbit Coupling Effects in Two-Dimensional Electron and Hole Systems (Springer Berlin Heidelberg, 2003).